\documentclass[preprint,authoryear,12pt]{elsarticle}

\usepackage{graphicx}
\usepackage{color}
\usepackage{amsmath,amsfonts,amssymb}
\usepackage{tabularx}
\usepackage{ulem} 

\journal{Safety Science}

\begin{document}

\begin{frontmatter}



\title{Modelling social identification and helping in evacuation simulation}

\author[a,b]{I. von Sivers\corref{cor1}} 
\author[c]{A. Templeton}
\author[b]{F. K\"unzner}
\author[a]{G. K\"{o}ster}
\author[c]{J. Drury}
\author[c]{A. Philippides}
\author[b]{T. Neckel}
\author[b]{H.-J. Bungartz}

\address[a]{Munich University of Applied Sciences, Lothstr. 64, 80335 Munich, Germany}
\address[b]{Technische Universit\"{a}t M\"{u}nchen, Boltzmannstr. 3, 85747 Garching, Germany}
\address[c]{University of Sussex, Falmer, Brighton, BN1 9QH, United Kingdom}


\begin{abstract}
Social scientists have criticised computer models of pedestrian streams for their treatment of psychological crowds as mere aggregations of individuals. Indeed most models for evacuation dynamics use analogies from physics
where pedestrians are considered as particles.
Although this ensures that the results of the simulation match important physical phenomena, such as the deceleration of the crowd with increasing density, social phenomena such as group processes are ignored. 
In particular, people in a crowd have social identities and share those social identities with the others in the crowd. The process of self categorisation determines norms within the crowd and influences how people will behave in evacuation situations.
We formulate the application of social identity in pedestrian simulation algorithmically. The goal is to examine whether it is possible to carry over the psychological model to computer models of pedestrian motion so that simulation results correspond to observations from crowd psychology. That is, we quantify and formalise empirical research on and verbal descriptions of the effect of group identity on behaviour. 
We use uncertainty quantification to analyse the model's behaviour when we vary crucial model parameters.
In this first approach we restrict ourselves to a specific scenario that was thoroughly investigated by crowd psychologists and where some quantitative data is available: the bombing and subsequent evacuation of a London underground tube carriage on July 7\textsuperscript{th} 2005.

\end{abstract}


\end{frontmatter}


\section{Introduction}  
The importance of evacuation simulations  for pedestrians  is generally accepted for designing buildings and for ensuring safety  at mass events.
Multiple models for pedestrian simulation were developed in the past decades. The forerunners and still most popular among them are force based models \citep{helbing-1995,langston-2006,dietrich-2014} and cellular automata \citep{gipps-1985,blue-1997,schadschneider-2001,kirik-2007},
but several alternatives were added to the portfolio in more recent years \citep{dijkstra-2006,sud-2008,seitz-2012}.

While having a realistic locomotion model is an important basis for simulating pedestrian evacuations, many crucial aspects of social behaviour have been neglected. In particular, the emergence and effects of group behaviour have been examined extensively in empirical research by social psychologists but are often missing in computer models \citep{templeton-2015}.

On the other hand, there are a number of publications on pedestrian motion models which incorporate social and psychological behaviour \citep{pan-2007,chu-2011,tsai-2011,chu-2013}. Yet, these publications mostly describe computer frameworks 
designed to incorporate many possible but as yet unspecified behavioural models. That is, their focus is on the implementation of the framework. Notably, they do not attempt to emulate empirical findings from psychology. Furthermore, information on how the behaviours they implemented are modelled in detail is missing in most cases. Another problem is the large number of
parameters  used to control the interworking between the software modules that instantiate the supposed behavioural models.
These three aspects -- the distance of the models to the underlying theories, the lack of detailed modelling information, and the number of parameters to calibrate -- make it nearly impossible to replicate the models and to check the models against observations. From a safety scientist's point of view, this is a severe drawback.

\begin{figure}
	\centering
		\includegraphics[width=1.00\textwidth]{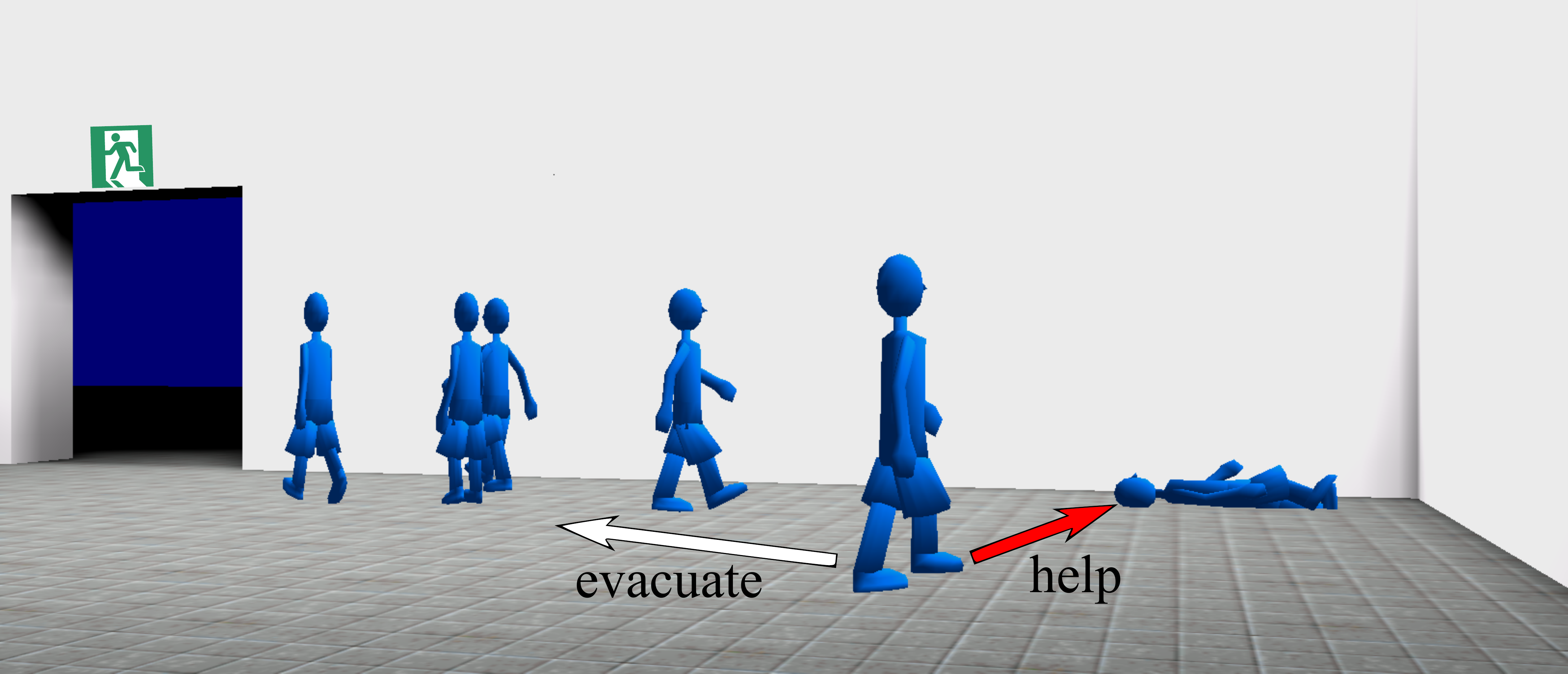}
		\caption{In an evacuation people have to decide whether to evacuate as quickly as possible or to
		help others. We explain psychological mechanisms behind these decisions, formalise them in
		a computer model and show which model parameters are decisive for safety science.}
	\label{fig:SIED_Exit}
\end{figure}

This is the gap that we attempt to close. With the Social Identity Model Application (SIMA) for pedestrian simulation, we present a step towards combining research from psychology and computer science. We focus on one pivotal behaviour that was observed in several evacuations: helping others (see Figure \ref{fig:SIED_Exit}). In modelling, we  directly follow the ideas of the self-categorisation theory (SCT) \citep{turner-1987} and the social identity theory (SIT) \citep{tajfel-1979} which are both part of the social identity approach. We describe how to algorithmically formulate helping behaviour in an evacuation and
how to choose the parameters so that other scientists can replicate, validate, and use the model. We keep the model independent of the locomotion level, and the parameter space small. We argue that the model is parsimonious without being reductionist and thus falsifiable \citep{seitz-submitted2015b}.

\section{Materials and methods}
This contribution combines findings from social psychology and computer science. To address an audience from both disciplines it is necessary to introduce the methods with a little more depth and detail than one would, perhaps, use in either one of the single disciplines. 

\subsection{Social identity theory and self-categorisation theory} 

\subsubsection{Incorporating evidence from crowd psychology}

Our model is based upon extensive empirical research on collective behaviour by social psychologists. There are numerous real life examples of collective behaviour where people act together as a group: for example orchestras, football fans, and sports teams. Two prominent theories which provide insight in to how this group behaviour emerges are social identity theory \citep{tajfel-1979} and self-categorisation theory \citep{turner-1987}. According to social identity theory, people have multiple social identities which are distinct from the identity of a person as an individual because they refer to one’s identity as part of a social group, such as a fan of a certain sports team. Self-categorization theory refers to the process whereby one categorises oneself as an individual or a group member. It suggests that collective behaviour occurs through the process of depersonalisation, where individuals self-stereotype themselves in line with their group. This occurs through a transformation of one's identity from the \textbf{personal} self to the \textbf{collective} self. It is this self-categorisation as a group member which makes collective behaviour possible. It  can therefore explain the behavioural differences between a \textbf{physical} crowd of individuals (who are simply in the same location together) and a \textbf{psychological} crowd (where people in a crowd act together).

The effect of social identities on people’s behaviour is crucial to understand for crowd modellers who aim to simulate psychological crowds. Research has shown that a shared social identity amongst crowd members increases the prevalence of supportive behaviours among people in emergency evacuations \citep{drury-2009b}.  For example, when they share a social identity people evacuating may be more likely to  coordinate  their  walking behaviours with others and  by letting  them move first rather than competing for the same exit. Research on emergency mass decontamination has also demonstrated that  social identity  is key to understanding  the coordination  of queuing behaviour,  showing that  that  members of the public are more likely to participate in queuing if they identify with the person organising the situation \citep{carter-2014}.

An example of collective  behaviour  in an emergency evacuation comes from the July 7\textsuperscript{th} London bombings \citep{drury-2009c}. In this paper, we will focus upon on this event  which has been analysed  by social psychologists.

\subsubsection{The event: the London bombings, 7\textsuperscript{th} July 2005}

At 8.50am, during the peak rush hour in central London, three bombs were set off simultaneously in the London underground. The bombs were coordinated so that they detonated on three separate tube lines when the tube trains were between busy stations. The passengers in the tube trains were plunged into darkness and could not know if there were going to be further explosions, with no information when help would arrive. Emergency services did not reach them for some time. Over 700 people were injured in the attacks and 52 people were killed. In this emergency situation, the survivors of the bombings came together to tend to the injured and find a way of safely evacuating. In contrast to portrayals of crowds as panicking and acting selfishly to evacuate, research has shown that the opposite occurred. In the aftermath of the disaster, \cite{drury-2009c} collected 141 contemporaneous accounts from survivors. 140 of the survivors reported seeing helping behaviour and mutual aid, such as offering water to others, providing first aid, and applying makeshift bandages. Only three of the survivors reported witnessing selfish behaviour. This was replicated in internet sources and the public enquiry; first-hand accounts showed that 42 people out of 127 described seeing help and only 11 described seeing people act selfishly.

Crucially, \cite{drury-2009c} found that survivors reported feeling part of a group with the other survivors. Many participants gave accounts of where people coordinated to help the group to escape. In fact, more survivors described the behaviour of the crowd as calm and orderly, rather than as panicking given the unexpected emergency which they were faced with. An important way that the survivors assisted safe evacuation was by providing mutual aid through cooperating and coordinating with each other. There was evidence of orderly queuing and people allowing others to go first, which in turn  helped people overall by making the evacuation more safe. Other examples include leaving in a calm manner rather than rushing  or pushing  past  each other which could have caused hindrance to the evacuation of the group overall.

\cite{drury-2009c} demonstrate how survivors emotionally supported one another, allowed others to evacuate first, and stayed behind with people (who were previously strangers) at a personal risk to themselves. This behaviour was relatively common across the survivors, and when combined with the reports of feeling as part of a group it is in line with the idea that the commuters shared a social identity which was invoked through the common fate of the emergency situation. Although there are numerous examples of this collective mutual aid in emergency situations, in this paper we will focus on one key helping behaviour that was documented in the London bombings: assisting injured people to evacuate safely. We argue that in order to adequately simulate collective behaviour in emergency and disaster events, modellers should heed the research conducted by social psychologists on crowd behaviour. Specifically, modellers should focus on the role of a shared group identity and incidents of ingroup helping by incorporating aspects of self-categorisation theory into their simulations. As such, we use the London bombings scenario to propose a realistic model of collective behaviour which combines the comprehensively validated Optimal Steps Model for locomotion with principles from the well-established social identity theory and self-categorisation theory.

\subsubsection{Collecting empirical evidence}

The helping behaviour that is modelled in SIMA is taken from accounts given by survivors of the July 7\textsuperscript{th} 2005 London bombings and the behaviour of the crowd in the aftermath. We have based this on the research by \cite{drury-2009c} which examined survivors' experiences of the bombings. They collated and analysed the survivor's perceptions of people's behaviour and the feelings that they experienced during the event. This was conducted by collecting one hundred and forty one accounts in contemporaneous newspaper material, in addition to personal archives and accounts from eighty one survivors which were recorded on the day or in the immediate aftermath of the bombings. Crucially, from this sample of survivors, they were asked about the level of danger that they felt they were in, how the others in the crowd behaved, whether people performed helping behaviour or acted selfishly, and whether their perception towards others changed throughout the course of the event (for a comprehensive account of the methodology, see \citep{drury-2009c}). Notably, support for these findings can be seen in \cite{drury-2009} which examines the accounts of crowd behaviour given by 21 survivors of 11 emergencies.

\subsection{The underlying pedestrian motion model}

For the locomotion level of our simulation, we use the Optimal Steps Model (OSM) \citep{seitz-2012,sivers-2013,sivers-2015}. 
As in many models, pedestrians are represented by circles, with radius 20 cm, that represent the
solid body. The model deviates from older approaches in its treatment of motion: As in
reality, pedestrians make steps to move forward.
They do not glide along smooth trajectories that resemble imaginary rails as in force-based models
or hop from cell to cell as in cellular automata. For this, each agent searches for the  possible
next position within a disk of which the radius is the agent's maximum stride length. The maximum
stride length is determined individually according to empirical findings that link the stride length
to free-flow speed \citep{grieve-1966,kirtley-1985,jelic-2012b,seitz-2012}. The free-flow speed,
that is, the speed an agent is supposed to prefer when walking uninhibited on a flat
surface, is a standard input parameter in pedestrian simulations.

The search for the next position constitutes a two-dimensional optimisation problem on the step disk.
The objective function is a superposition of dedicated utility functions that express closeness to the target, or rather a short travel time to the target \citep{koster-2014b}, sufficient interpersonal distance to other pedestrians \citep{hall-1966, sivers-2015} and sufficient distance to  obstacles \citep{seitz-2015}. The shorter the remaining travel time to the target the higher the utility becomes. The correct travel time, while skirting obstacles that hide the target from direct view, is expressed through the solution of a mathematical equation: the eikonal equation \citep{kretz-2009,kretz-2010,hartmann-2010}. Originally, the eikonal equation describes the arrival time of a wave front that spreads out from an area, in our case, the target. The arrival time of this imaginary wave front
coincides with the remaining travelling time of the agents in the model. All local utility or cost functions are chosen on a compact support, that is, they are truly zero at a suitable distance. This avoids numerical cut-off errors.  
The update scheme for the pedestrians is event driven \citep{seitz-2014b}. That is, each agent steps ahead according to its individual  stepping frequency which is predefined from the agent's free-flow speed and maximum stride length \citep{seitz-2012, seitz-2015}. Each step is an event in the event queue and occurs at its ``natural'' time. 

To make replication possible, the parameters for the locomotion model are compiled in Table \ref{tbl:parametersMotion}. 
The Optimal Steps model is embedded into the VADERE simulation framework at Munich University of Applied Sciences. Each module
of the framework is verified using an automatic test suite. The model itself has been extensively
validated \citep{seitz-2012, sivers-2013b, dietrich-2014, seitz-2014b, koster-2014b, sivers-2015,
koster-2015, koster-2015b} including independent implementations \citep{kneidl-2015}. Thus it is  a
reliable basis on which to build modules that instantiate social behaviour.

\begin{table}
	 \centering
		\begin{tabularx}{0.95\textwidth}{|p{0.1\textwidth}|X|l|}
			\hline
			Param. & Description & Value \\
			\hline
			$\delta_{int}$ & intimate distance 					& 0.45 $m$ \\
			$\delta_{pers}$ & personal distance 				& 1.20 $m$ \\
			$\delta_o$ & distance kept from obstacles 	& 0.8 $m$ \\
			$\mu_p$ & strength of `pedestrian avoidance'					 & 5.0 \\
			$a_p$ & moderation between intimate and personal space & 1.0 \\
			$b_p$ & transition between intimate and personal space & 1.0 \\
			$\mu_o$ & strength of `obstacle avoidance' & 6.0 \\
			$v_{f}$ & mean free-flow speed						 & 1.6 $\frac{m}{s}$ \\
			$\sigma_{f}$ & variance of the free-flow speed						 & 0.26 $\frac{m}{s}$ \\
			
			\hline
		\end{tabularx}
		\caption{The parameters for the OSM.}
     \label{tbl:parametersMotion}
\end{table}

\subsection{The theory of uncertainty quantification}

The field of uncertainty quantification (UQ) addresses the problem that not all parameter values in
the simulation are exactly determined: they are uncertain. Uncertainty quantification assumes distributions of the parameters and, thereby, allows
to analyse their impact on the solution of the simulation. Probability density functions (PDFs) may
be provided directly or statistical moments may be specified.

In recent decades, several methods beyond traditional Monte Carlo simulations have been developed. For a general overview see \citep{smith-2014} and \citep{xiu-2010}.
Uncertainty quantification distinguishes between forward and inverse uncertainty quantification
simulation.
The former analyses the impact of the uncertain parameters on the model
whereas the latter tries to determine the distribution of the input parameters.
In this work, we will use forward uncertainty quantification. 
Also we will employ non-intrusive methods where, in contrast to intrusive methods, the underlying model remains untouched.

\begin{figure}
  \centering
  \includegraphics[width=300px,viewport=0.5cm 5cm 24.5cm 16cm,clip]{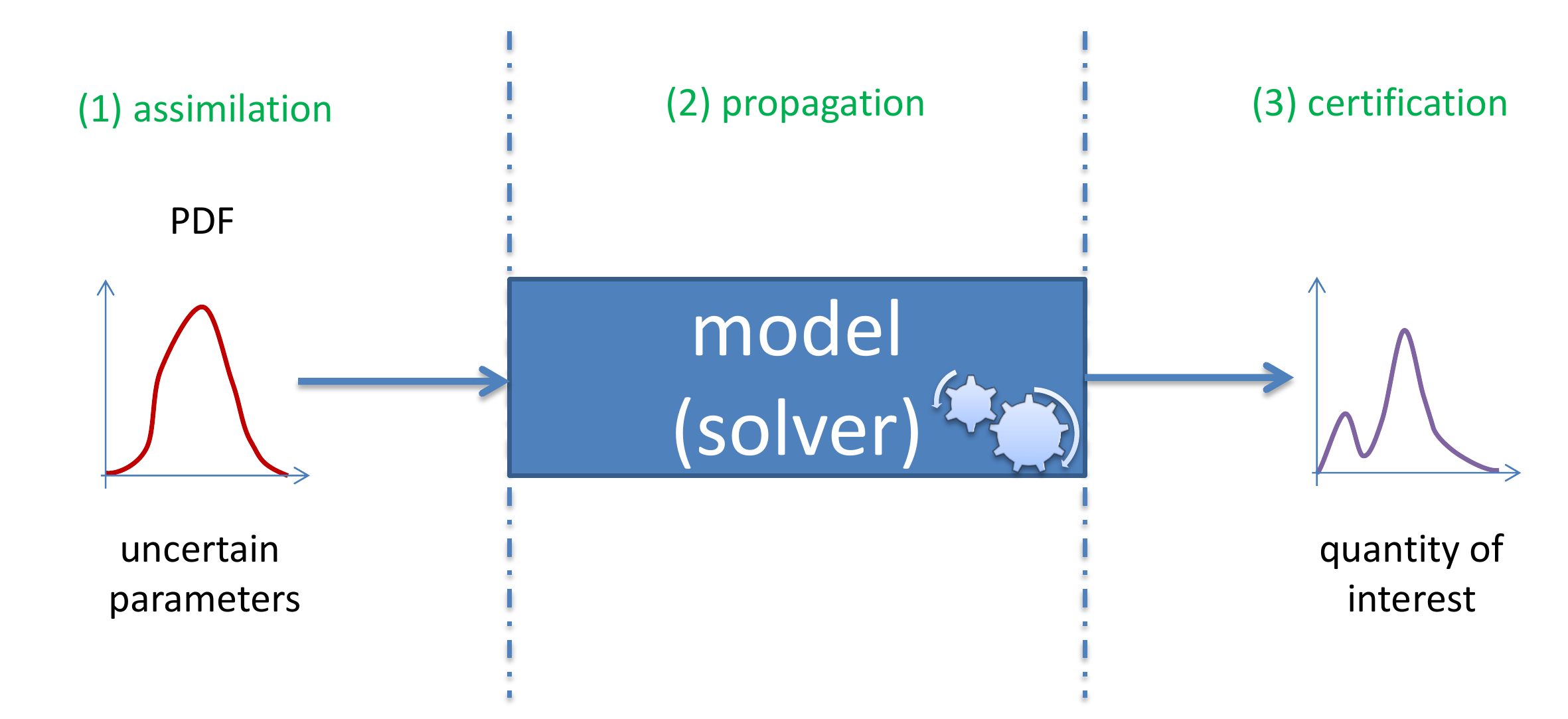}
  \caption{Illustration of the forward uncertainty quantification phases: assimilation,
  propagation, and certification.}
  \label{fig:UQ_Phases}
\end{figure}

A forward simulation in the context of uncertainty quantification contains three phases
\citep[p.17-7]{iaccarino-2008}: the assimilation phase, the propagation phase, and the certification
phase (Figure \ref{fig:UQ_Phases}).
In the assimilation phase, the uncertain parameters are defined and prepared with the help of
probability theory. For each uncertain parameter, a suitable probability distribution function (PDF)
has to be chosen.
In the propagation phase the model is evaluated. For this, the parameters are propagated through the model multiple times.
The number of propagations necessary to obtain  good results depends on the specific uncertainty quantification method.   
In the certification phase  the output of each propagation step is collected.  The impact of the
uncertain parameters on one or several quantities of interest is observed. Often, the mean and the variance of
an output quantity is calculated, but higher statistical moments are also possible
\citep[p.67]{xiu-2010}. Typical quantities of interests (QoI) are physical, domain, or timing
values.

Here, we focus on parametric uncertainty: We want to understand the impact of uncertain parameters, such as the number of injured persons, on the simulation results. We choose stochastic collocation (SC) with the pseudospectral approach \citep{xiu-2007} as uncertainty quantification method. This
fits our requirements because the method is non-intrusive, suitable for the analysis of models with
few parameters, and needs less computational effort than other methods \citep{xiu-2009}.

Stochastic collocation with the pseudospectral approach (SC-GPC) is based on the generalised
polynomial chaos (GPC) \citep[p.57-67]{xiu-2010} expansion. The idea is to separate the
spatio-temporal part of a random process $U(x,t,\boldsymbol{\zeta})$  from
the purely random part (see \citep[p.209]{smith-2014}). In our case, $U(x,t,\boldsymbol{\zeta})$ is the
theoretical `analytical' solution of the considered pedestrian evacuation scenario.

The generalised polynomial chaos expansion for $U(x,t,\boldsymbol{\zeta})$ is defined as
\citep[p.143]{xiu-2003}
\begin{equation}
    \label{eq:gPC_inf}
    U(x,t,\boldsymbol{\zeta}) = 
        \sum_{j=0}^\infty 
            \underbrace{ c_j(x,t)}_{\substack{\text{spatio-}\\\text{temporal}}} 
            \cdot
            \underbrace{ \Phi_j(\boldsymbol{\zeta}) }_{\text{random}},
\end{equation}
which depends on space $x$, time $t$, and the vector of random parameters $\boldsymbol{\zeta} =
(\zeta_1, \ldots, \zeta_M)$ with $M$ independent random variables. $c_j(x,t)$
are the coefficients of the expansion representing the spatio-temporal part and are denoted by
$c_j$ for the sake of simplicity in the following.
Their computation depends on the specific generalised polynomial chaos method
\citep[p.209]{smith-2014} one uses.
The functions $\Phi_j$ are orthogonal polynomial basis functions that correspond to the
distributions of the random variables. To evaluate Equation \eqref{eq:gPC_inf} numerically, the
expansion is truncated after N terms.

In the stochastic collocation with the pseudospectral approach, the model is evaluated on so-called
collocation points. The coefficients $c_j$ of the generalised polynomial chaos expansion are
approximated by applying a projection with an integration rule: Once the uncertain parameters are
chosen, the collocation points \citep[p.299] {xiu-2007} $z_i$ and weights $w_i$ ($i = 1$, \ldots,
$Q$) for the integration rule can be generated (assimilation).
To propagate the uncertainty through the model and
to calculate the coefficients $c_j$, the model $u(z_i)$ (i.e.~our pedestrian simulation) and the
orthogonal basis functions $\Phi_j(z_i)$ have to be evaluated at the generated collocation points
$z_i$.

Finally, the quantities of interest are extracted (certification). In our example, we have
one quantity of interest: the number of pedestrians that have not yet reached safety in an evacuation
scenario resembling the London bombings in 2005. To extract the mean $\mu_{u_{N}}$ and the variance
$\sigma^2$ of the desired quantity of interest, we can directly use the coefficients $c_j$. For the
mean, one has to evaluate \cite[p. 67]{xiu-2010}

\begin{equation}
\label{eq:sc_mean}
    \begin{split}
        \mu_{u_{N}} 
            \approx 
           &\underbrace{ \sum_{i=1}^Q u(z_i)\Phi_0(z_i)w_i }_{ c_0 }
           = c_0,
    \end{split}
\end{equation}

and for the variance $\sigma^2$ \cite[p. 67]{xiu-2010}

\begin{equation}
\label{eq:sc_variance}
    \begin{split}
        \sigma^2  
           \approx
           &\sum_{j=1}^N (\underbrace{\sum_{i=1}^Q u(z_i)\Phi_j(z_i)w_i }_{ c_j })^2
           = \sum_{j=1}^N c_j^2,
    \end{split}
\end{equation}

with $\sigma$ being the standard deviation.

We implemented a Python program using the chaospy library \citep{feinberg-2015} to
realise the stochastic collocation with the pseudospectral approach (for results, see
Section \ref{sec:uq_results}).
Since we are going to use uniform distributions for the computations (see Section
\ref{sec:uq_setup}), Legendre polynomials are used as orthogonal basis functions $\Phi$ (according
to \citep[p.626]{xiu-2002}). We choose the Gauss quadrature as integration rule, and the order of
the polynomial chaos expansion is fixed to $N=6$. 
The number of required simulation runs depends on the number of collocation points and the
number of uncertain parameters. In our example, we use $Q=21$ collocation points for each uncertain
parameter.
Hence, for the settings with one uncertain parameter, 21 simulation runs are required, and for the
setting with three uncertain parameters 9126 ($=21^3$).

\section{Results: Formalisation of social identity and helping}

In this section, we present the way we formalise the self-categorisation process and the
resulting helping behaviour. We explain the structure of our model through flow diagrams and declare the parameters.
We build our model on evidence from empirical research that shows the effect of social identity on behaviour. Since the social psychological models are very complex and detailed we must decide which parts to take into account and which to drop to construct a model that remains falsifiable.

Self-categorisation theory suggests that, when people categorise themselves as being in the same group, they are more likely to support each other \citep{levine-2005}. For the simulation of emergency evacuations, this basic idea is essential. Other parts of the self-categorisation theory that are important to social psychologists -- for example a changing degree of social identity during the evacuation -- have to be postponed until there is data to substantiate the mechanisms of the change.
In the next four subsections, we algorithmically formulate shared social identity and its effect on helping behaviour. 

\subsection{Social Identity Model Application}

We want to enable independent researchers to use the new social model, the \textbf{Social Identity Model Application (SIMA)}  with any locomotion model and in any simulation framework. The Optimal Steps Model and the VADERE simulation framework of this contribution are merely examples among several well validated choices.
We achieve our goal by defining an interface that exclusively consists of the target and the velocity of the pedestrians, the two typical input parameters  for a locomotion step. 
The Social Identity Model Application is called before the execution of the locomotion module in every time step of the simulation. The outcomes of the SIMA call are adjusted targets and velocities of the agents.  

\begin{figure}
	\centering
		\includegraphics[width=\textwidth]{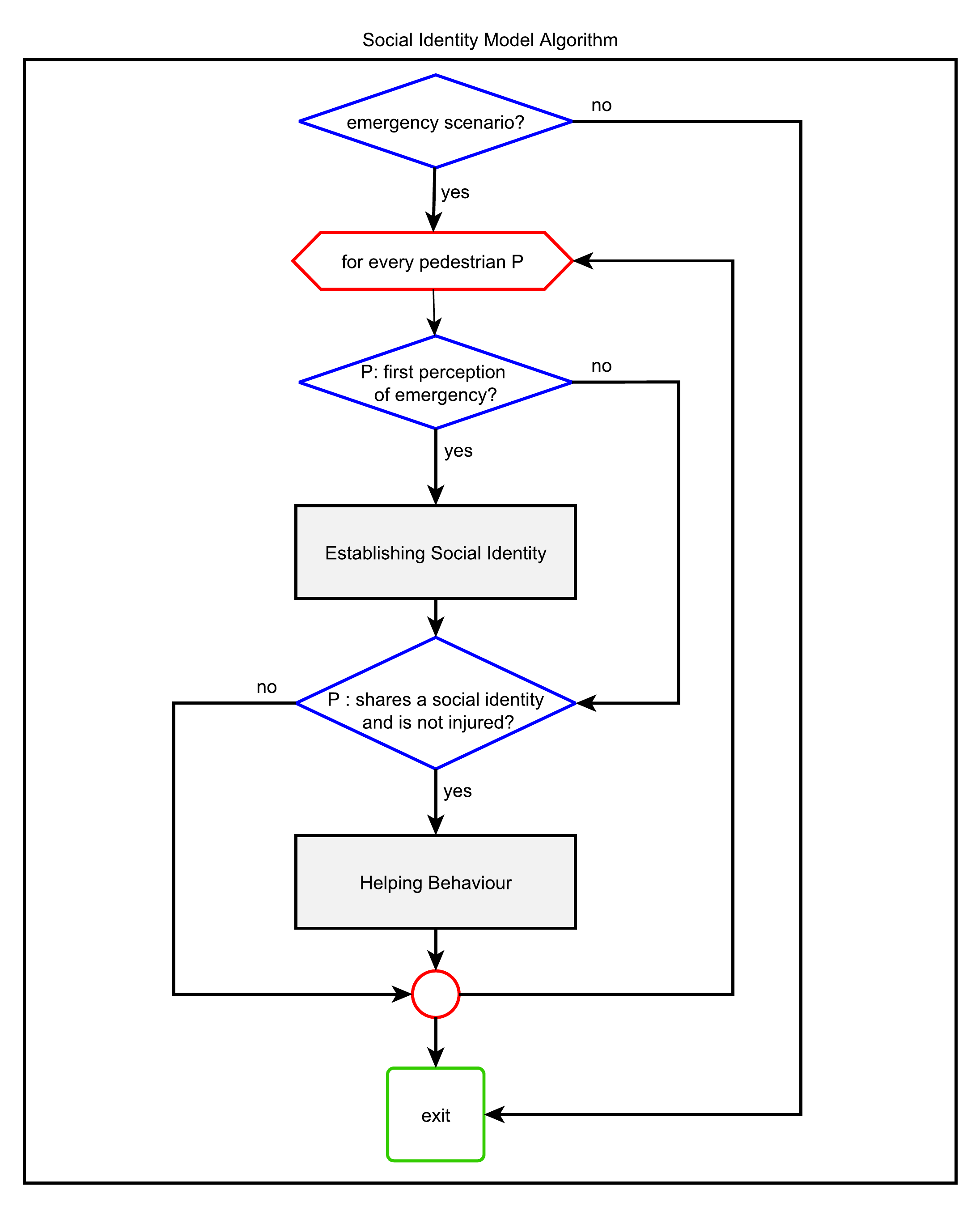}
	\caption{General flow chart of the Social Identity Model Application.}
	\label{fig:SIMA_general}
\end{figure}

The Social Identity Model Application consists of two main components: the social identity component (\textbf{Establishing Social Identity}) described in Section \ref{subsec:EstablishingSocialIdentity}
and the helping behaviour component (\textbf{Helping Behaviour}) described in Section \ref{subsec:HelpingBehaviour}. Figure \ref{fig:SIMA_general} shows the main loop of the Social Identity Model Application with these two key components. A new type of agents, badly injured pedestrians, is introduced (see Sec. \ref{subsec:Badlyinjuredpedestrians}).   

The \textbf{Establishing Social Identity} component is called the first time a pedestrian recognises
the emergency. This follows findings from social psychological research that people categorise themselves as ingroup members when faced with the common fate of an emergency.
The component \textbf{Helping Behaviour} is relevant during the whole duration of the simulation. It is called for pedestrians who share a social identity. Pedestrians who do not share a social identity head straight for safety, that is, they evacuate without caring for others.

\subsection{Establishing social identity}
\label{subsec:EstablishingSocialIdentity}

\begin{figure}
	\centering
		\includegraphics[width=\textwidth]{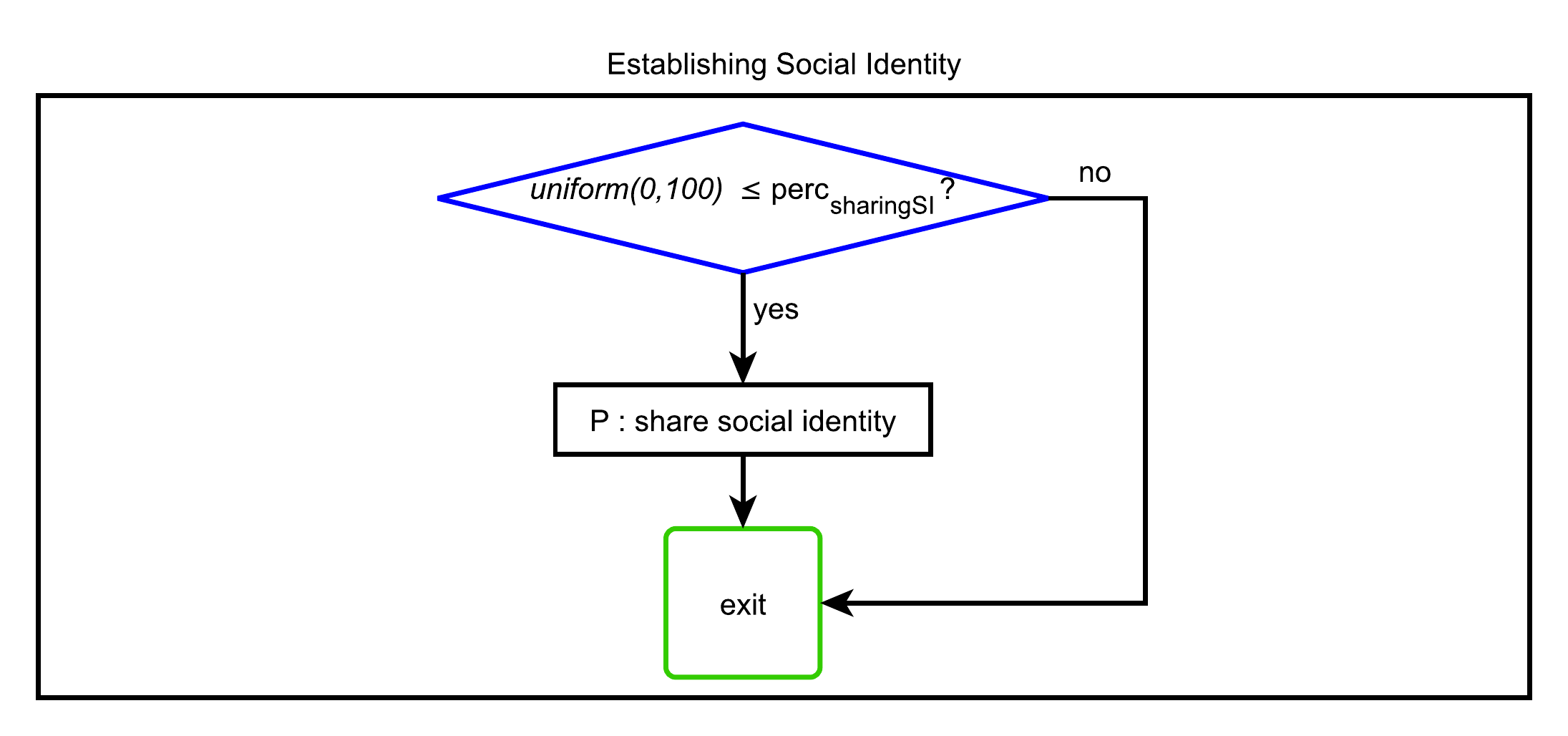}
	\caption{Flow chart of how the social identity for pedestrian $P$ is established at the beginning of the simulation. The percentage of pedestrians sharing a social identity is $perc_{sharingSI}$.}
	\label{fig:SIMA_ESI}
\end{figure}

The first step in the model is to establish the social identity. However, not every pedestrian in an emergency shares a social identity \citep{drury-2009}. Thus, we define a parameter $perc_{sharingSI}$. Pedestrians are randomly selected to share a social identity or not according to $perc_{sharingSI}$. The procedure for one pedestrian in the scenario is visualised in Figure \ref{fig:SIMA_ESI}.

\subsection{Badly injured pedestrians}
\label{subsec:Badlyinjuredpedestrians}

To model helping behaviour, we first need to introduce another type of agent: a badly injured pedestrian. In emergencies in general, people can get hurt by fire, bombs, plunging building parts or for other reasons. Although in the London bombings there were some pedestrians who suffered minor injuries, we neglect them here. Their behaviour could be easily modelled by reducing their speed. Instead, we focus on badly injured pedestrians who need assistance from others to evacuate. 

A pivotal element in evacuation models is the target that each agent moves towards. Usually this is a `safe area'. We model the immobility of badly injured pedestrians by fixing their target at their current position. Thus they remain stationary.  
They cannot evacuate without assistance from unharmed pedestrians. As soon as such an aide arrives we turn the aide into the injured agent's target. By doing this, the injured agent becomes dependent on the aide. The aide's target is set to the `safe area' and the pair evacuate at a 
reduced speed $v_{inj}$. This helping behaviour matches reports by survivors \citep{baker-2002,johnson-2005,tucker-2007}. 

\subsection{Helping behaviour}
\label{subsec:HelpingBehaviour}

\begin{figure}
	\centering
		\includegraphics[width=\textwidth]{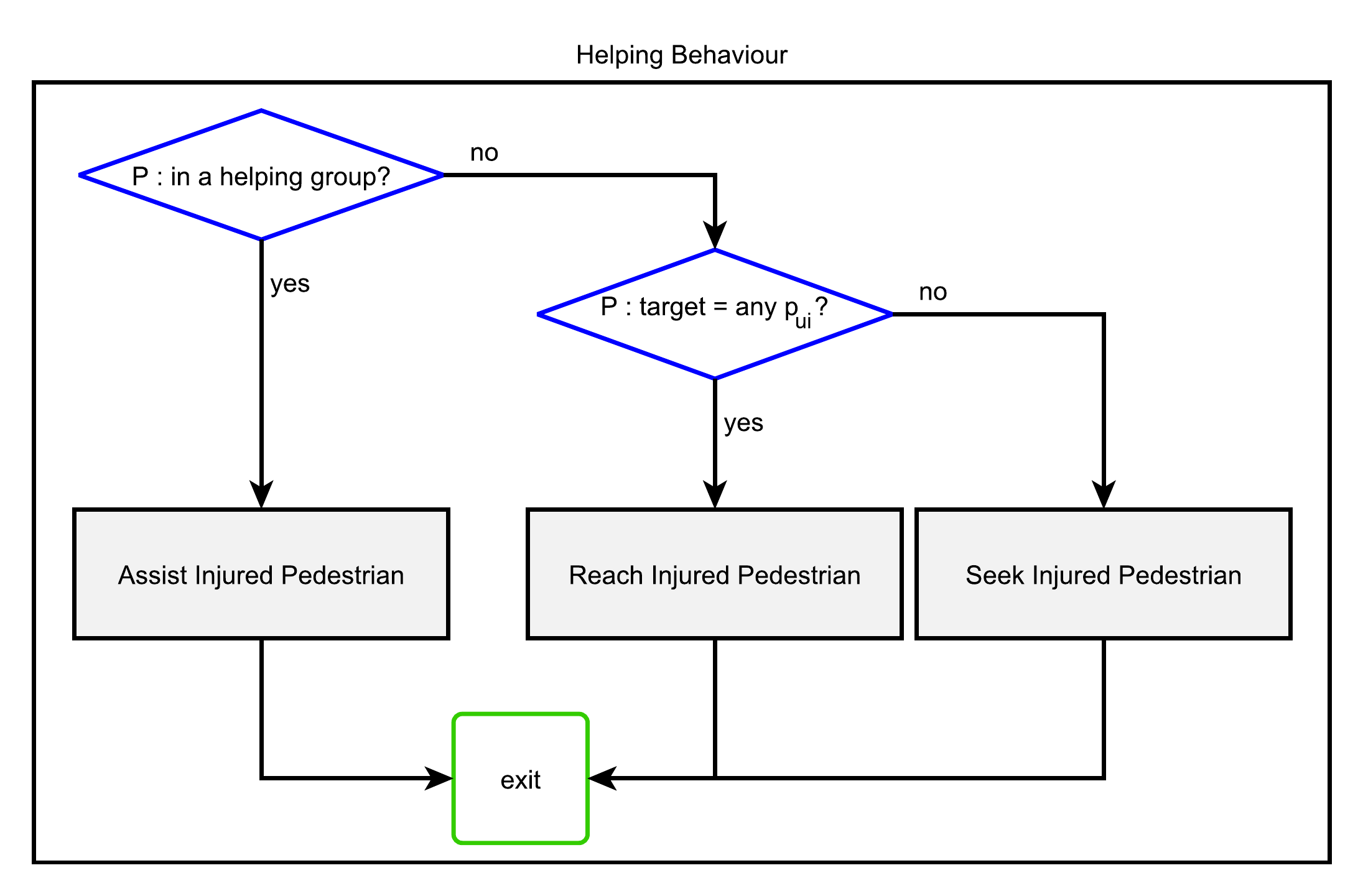}
	\caption{Flow chart of the helping behaviour for pedestrian $P$ during the emergency; $\mathrm{p_{ui}}$ is the abbreviation for unaided and injured pedestrian.}
	\label{fig:SIMA_HB}
\end{figure}

Reports on help between strangers and survivors in evacuations usually lack a description of what exactly people do to help.
However, one can imagine the typical steps that must be taken. Initially, an injured pedestrian needs to be detected by an aide. Subsequently, this aide must approach the injured pedestrian to finally physically support the injured person
while evacuating together. These assumptions are what we implement as helping behaviour (see Figure \ref{fig:SIMA_HB}). 

\begin{figure}
	\centering
		\includegraphics[width=\textwidth]{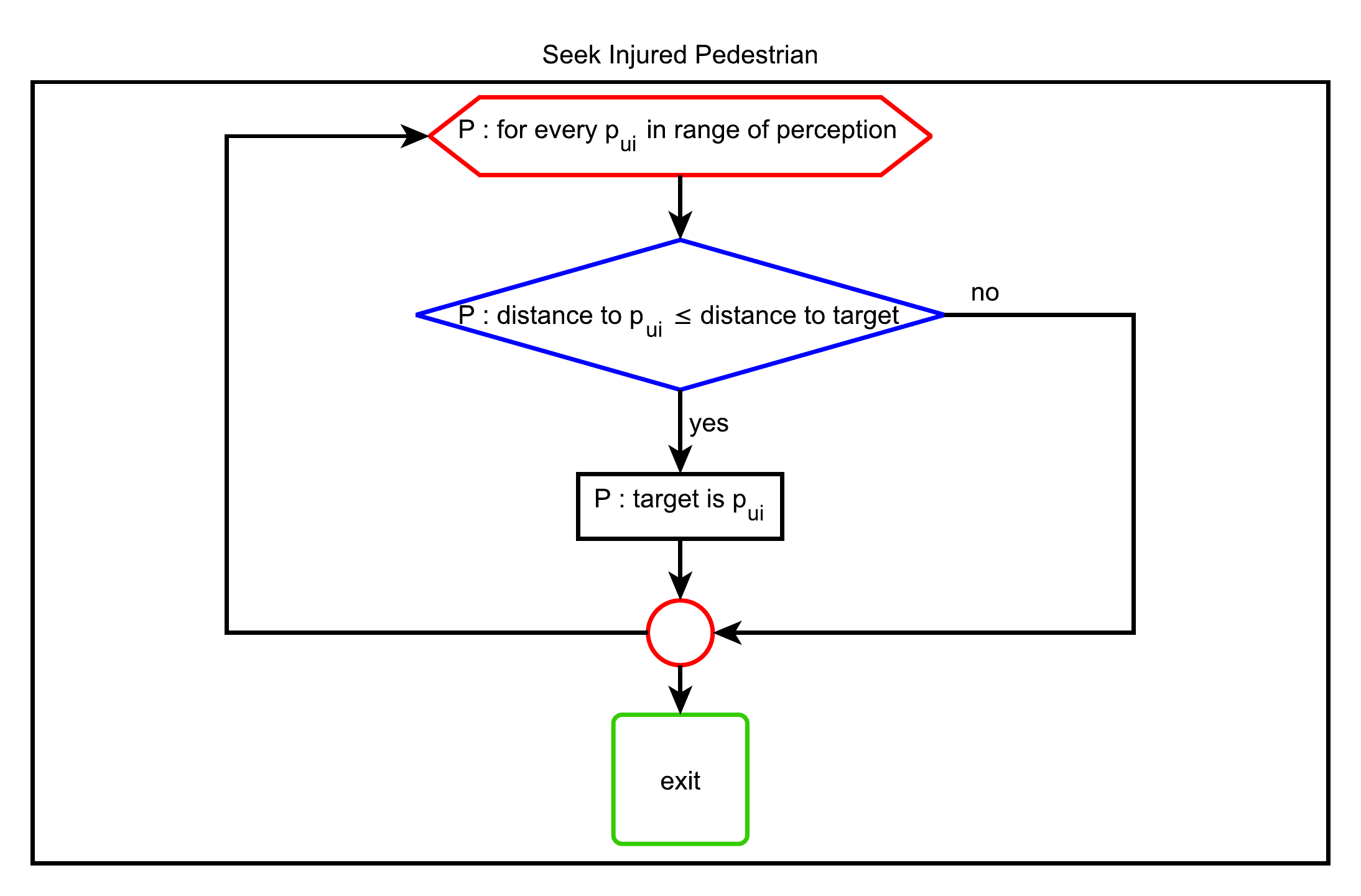}
	\caption{Flow chart of the helping behaviour when pedestrian $P$ looks for injured pedestrians and chooses whom to help. $\mathrm{p_{ui}}$ is the abbreviation for unaided and injured pedestrian. Target means the target of the pedestrian $P$. The range of perception is a radius of 10 m around the pedestrian.}
	\label{fig:SIMA_SIP}
\end{figure}

We assume the pedestrians choose the nearest injured and unaided pedestrian as the person to assist. The sub-component \textbf{Seek Injured Pedestrian} realises this searching behaviour. 
Each pedestrian has a range of perception that, in reality, depends on  the scenario or on the pedestrian's abilities. In our simulations, we simplify these dependencies by assuming that each pedestrians is aware of casualties 
within a radius of 10 m, neglecting visual obstructions or the fact that the train has different compartments and cars. We argue, that even through visual obstructions the shouts of injured people can be heard so the range of perception is far bigger than the range of vision. Each potential helper chooses the nearest unaided and injured pedestrian as new target (see Figure \ref{fig:SIMA_SIP}) and approaches this target. If there is no injured pedestrian in the range of perception the pedestrian heads to the `safe area'. 

\begin{figure}
	\centering
		\includegraphics[width=\textwidth]{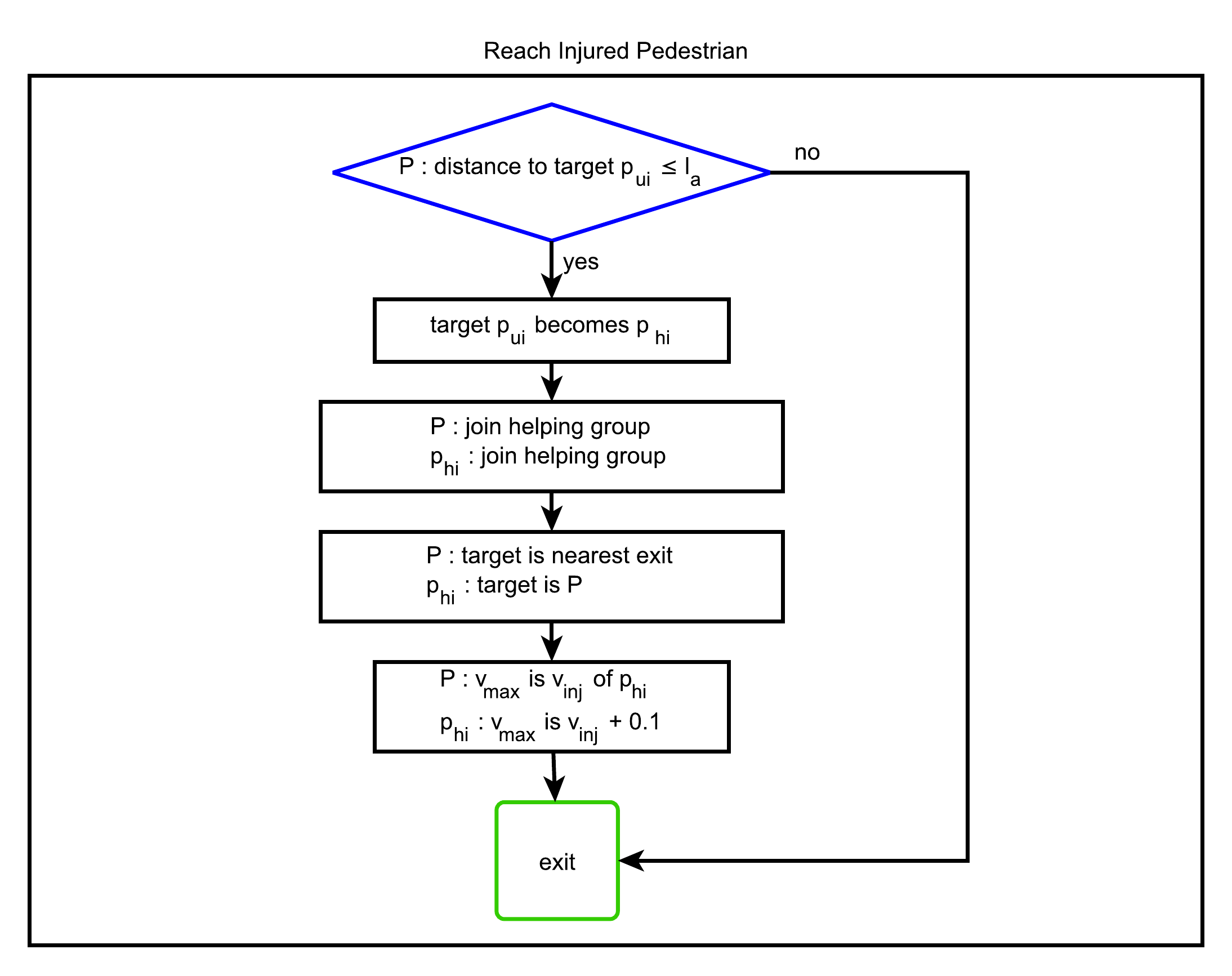}
	\caption{Flow chart of the helping behaviour when an injured pedestrian is reached. Parameter $l_a$ is the arm length of a pedestrian, $\mathrm{p_{ui}}$ stands for an unaided injured pedestrian, $\mathrm{p_{hi}}$ for helped injured pedestrian.}
	\label{fig:SIMA_RIP}
\end{figure}

The sub-component \textbf{Reached Injured Pedestrian} controls the situation when a pedestrian reaches the unaided and injured pedestrian. If there is less than an arm length between them they form are group and evacuate together. They both change status and targets. First, the unhelped injured pedestrian $p_{ui}$ becomes a helped injured pedestrian $p_{hi}$. 
This status change guarantees that other potential aides stop approaching and 
search for other casualties (see Figure \ref{fig:SIMA_HB}). Then, the aide selects the safe location as next target and reduces the free-flow speed to $v_{inj}$. Since the aide's location is set as target for the injured pedestrian, the latter automatically follows. The injured agent's free-flow speed is set slightly higher than $v_{inj}$ so that it does not fall behind or loose the aide.
The arm length $l_a$ is set to 60~cm. The single steps of the procedure are shown in Figure \ref{fig:SIMA_RIP}.

The last step of the helping behaviour is the sub-component \textbf{Assist Injured Pedestrian}. An agent in this stage evacuates with the injured agent without looking for other casualties. Notably, agents can only help one injured  throughout the simulation. From then on, both aide and injured pedestrians keep their respective targets and their reduced speeds. 
Note that, while the Social Identity Model Application can be applied for almost any locomotion model,  the concrete simulation outcomes, such as evacuation times, will most likely vary. 

\section{Results: quantification of uncertainty in the model}

In this section, we present simulation results for two scenarios, both inspired by the situation in the London bombings. In the first scenario, we demonstrate how the inclusion of helping behaviour affects the  evacuation of a whole tube train. The second scenario focuses on one carriage of the train. With this scenario, we analyse the impact of parameter variations on the simulation using  uncertainty quantification.

\subsection{Simulation of a whole train evacuation}
\label{subsec:simulation_whole_train}

Initially, we focus on the introduction of helping behaviour in a train evacuation scenario of a correctly dimensioned London Underground C69/C77 Circle Line train as in the July 7\textsuperscript{th} London bombings. The evacuation route, which in reality was a path along the tracks of the trains, is modelled as a long and narrow corridor that leaves no room for walking abreast. 
As during the real bombings, the scenario takes place during rush hour, so that every seat of the 192 seats in the train is occupied, that is, the initial positions of the
agents are given by the seat locations. For simplicity, we assume that the standing room is empty. 

The simulation parameters of the locomotion model, the Optimal Steps Model, are compiled in Table \ref{tbl:parametersMotion}. All parameters of the Social Identity Model Application for this simulation are compiled in Table \ref{tbl:parametersSocial}. In the absence of measured behavioural evidence for this scenario, we chose plausible values for the number of pedestrians who share a social identity and for the number of injured pedestrians. The speed of a person assisting a casualty is set to 0.6$m/s$. This value corresponds to observed slowest speeds of pedestrians with a walking handicap \citep{perry-1992}.

\begin{table}
	 \centering
		\begin{tabularx}{0.95\textwidth}{|p{0.2\textwidth}|X|c|}
			\hline
			Param. & Description & Value \\
			\hline
			$perc_{sharingSI}$ 	& people sharing a social identity 							& 0.8 \\
			$perc_{injPeds}$ 		& percentage of injured pedestrians 						& 0.1 \\
			$v_{inj}$ 					& speed of a helper with an injured pedestrian 	& 0.6 $\frac{m}{s}$\\
			\hline
		\end{tabularx}
		\caption{Parameters for the Social Identity Model Application.}
     \label{tbl:parametersSocial}
\end{table}

\begin{figure}
	\centering
		\includegraphics[width=\textwidth]{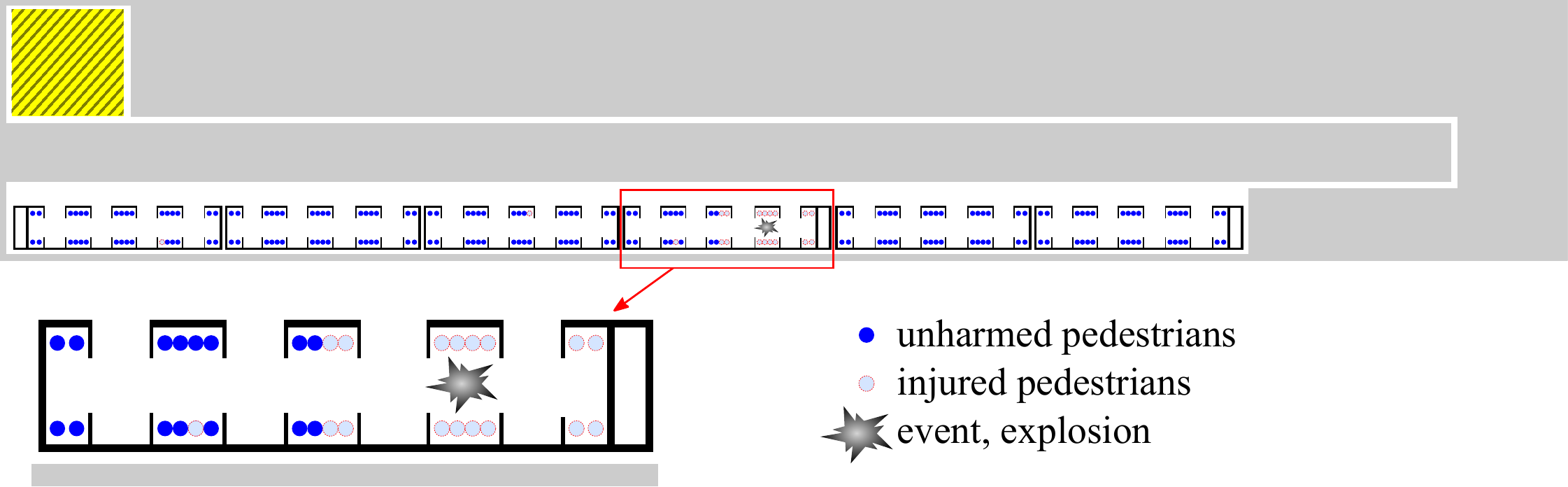}
	\caption{Evacuation scenario of a train. We assume that every seat is occupied (blue circles for unharmed passengers and red-rimmed light blue circles for injured passengers), but nobody is standing. Black lines indicate the partitioning inside the train and the walls. Light grey areas cannot be stepped upon. The escape route is the narrow white corridor that leads to safety. The safe area, and target, is indicated by the large (yellow, striped) rectangle in the upper left corner. 16 of the 19 injured pedestrians  are placed near the event (grey star), three more randomly in the train.}
	\label{fig:scenario_large}
\end{figure}

We assume that the bomb detonated in the third car and that, as a consequence, 19 pedestrians are badly injured: 16 casualties are near the bomb and 3 casualties are randomly positioned at other places. Among the remaining pedestrians, those who share a social identity are randomly chosen, according to the percentage set for the simulation run. See Figure~\ref{fig:scenario_large} for an illustration of the setting. 

\begin{figure}
	\centering
		\includegraphics[width=\textwidth]{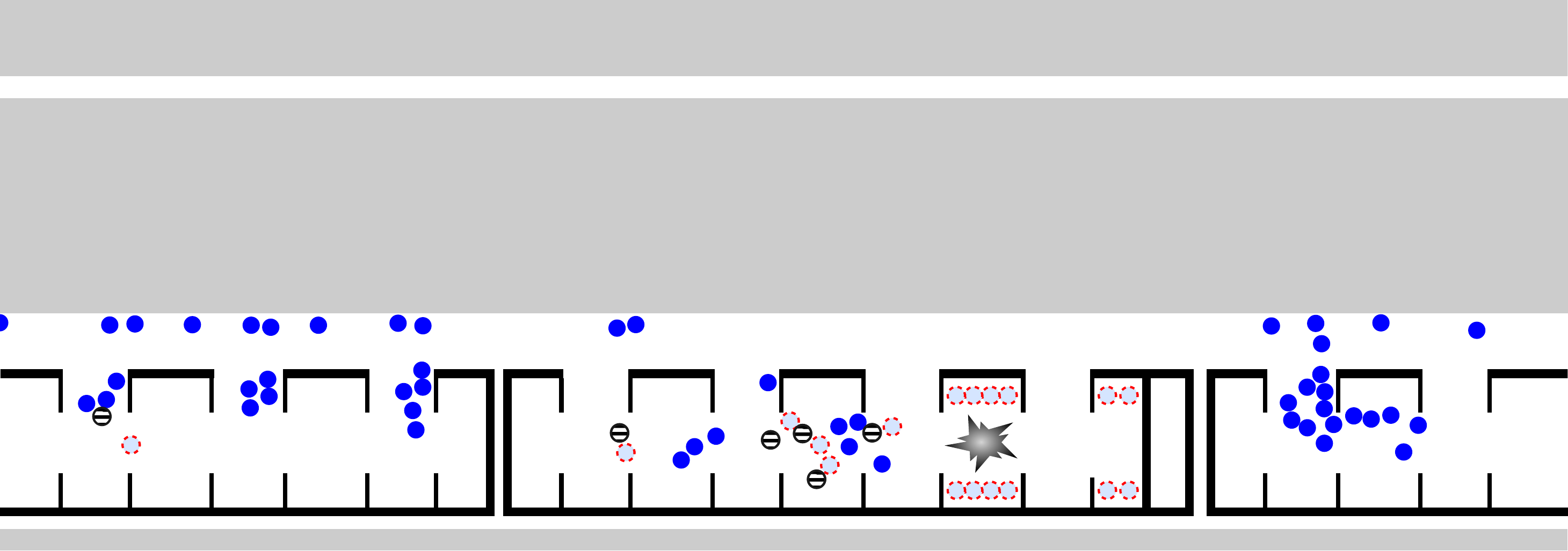}
	\caption{Evacuation scenario of a train. Close-up after the first seconds of the evacuation. First pairs of injured pedestrians (red-rimmed light blue cirlces) with helpers (black striped circles) form.}
	\label{fig:forming_groups}
\end{figure}

\begin{figure}
	\centering
		\includegraphics[width=\textwidth]{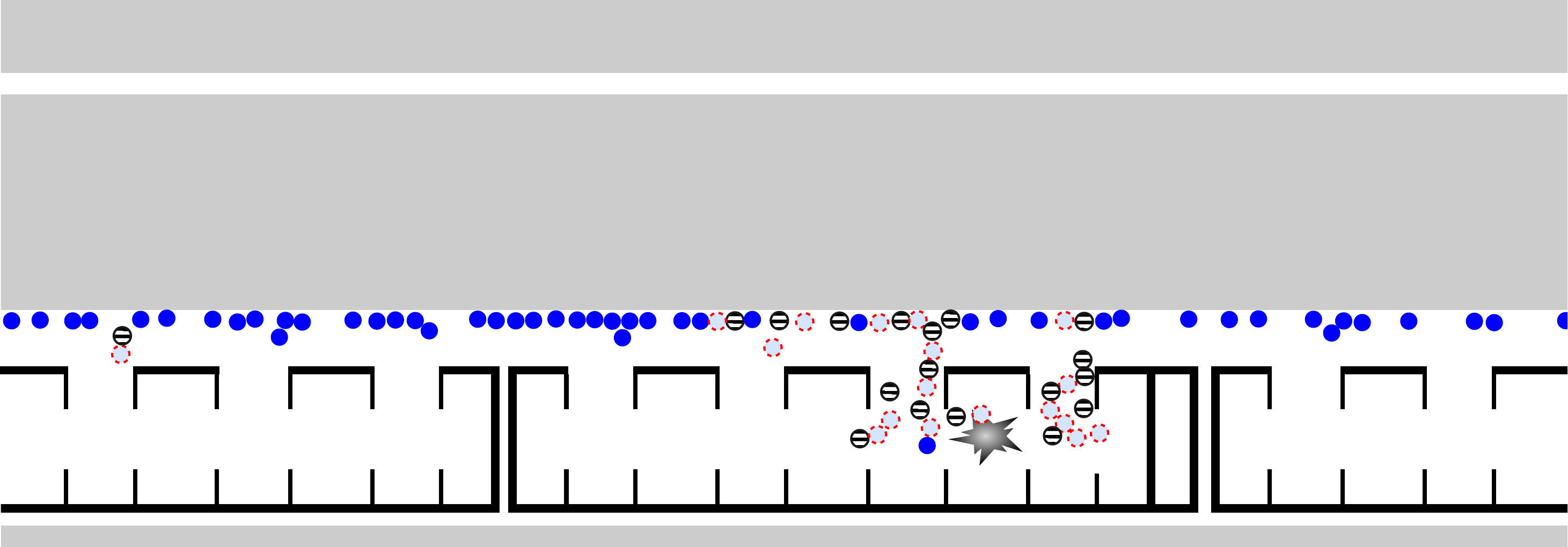}
	\caption{Evacuation scenario of a train. Close-up  after all pairs of aides and charges have formed. The unharmed pedestrians (blue circles) leave the train faster, the helpers (black striped circles) with injured  (red-rimmed light blue circles) are slower.}
	\label{fig:evac_normal}
\end{figure}

After a few seconds of the simulation, the first pairs of aides and injured pedestrians form. In Figure~\ref{fig:forming_groups} the helpers are depicted by black striped circles and the injured by light blue circles. At this point, some of the injured pedestrians are still without helpers. A few seconds later, all injured pedestrians are assisted (see Figure \ref{fig:evac_normal}). The other pedestrians (indicated by blue circles) evacuate on their own. In this model,  those pedestrians who do not help anybody leave faster than  those assisting injured pedestrians. This is a result of the reduced speed of the aides and injured pedestrians.

\begin{figure}
	\centering
		\includegraphics[width=\textwidth]{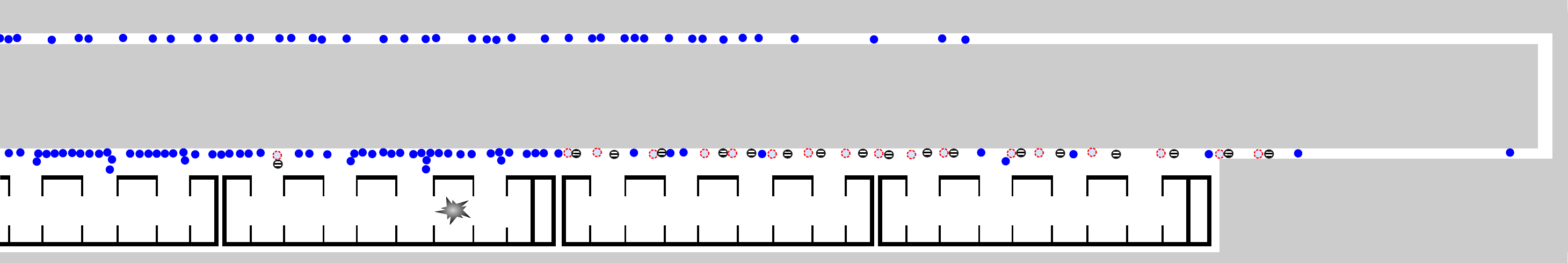}
	\caption{Evacuation scenario of a train. The slow pairs of helpers and charges (red-rimmed light blue and black striped circles) cause congestions.}
	\label{fig:jam_track}
\end{figure}

In a later state of the simulation, the escape route becomes congested. In this scenario, overtaking while walking along the evacuation path is not possible. Thus, faster agents get stuck behind aides with their charges.
See Figure \ref{fig:jam_track}.

All observations outlined in the simulation match accounts by survivors in emergencies \citep{johnson-2005,drury-2009c}. They reported that survivors assisted those who were injured before leaving and  formed orderly queues while evacuating.  
This is the behaviour that emerges in our simulations. We argue that  this constitutes  a qualitative validation  of the Social Identity Model Application. Quantitative validation  must be postponed  until suitable data  is available, possibly from evacuation drills or video footage
of future emergencies. For a longer discussion on validation challenges see \cite{sivers-2014b}.
In addition to qualitative validation,  we are able to provide statistical data 
on the variation of simulation outcomes using the techniques of  uncertainty quantification.

\subsection{Uncertainty quantification of the Social Identity Model Application}

In many applications, the precise value of an important parameter, such as the number of injured pedestrians in emergency planning, is unknown. Moreover, even if a measured value for the parameter is available, it is only correct within a margin of error. If the model is sensitive to the variation of this parameter the predictive power of the model is decreased.

 
For example, the new helping behaviour in the simulation has a crucial impact on the evacuation time. 
Clearly, the average evacuation time must depend on the number of injured pedestrians and helpers. In the extreme case, 
where at least one pedestrian is injured but nobody helps, the evacuation time is infinite. The same occurs
if everyone is injured. On the other hand, if there are no casualties then no time is
invested in searching for them and nobody is slowed down by helping. Evacuation is much
faster. A sample simulation with the setting from Section \ref{subsec:simulation_whole_train} with 10\% casualties results in an evacuation time of 467 seconds. In the same scenario without injuries, the evacuation time is 231 seconds. Only 10\% of casualties in this scenario doubles the evacuation time.

As long as there are uncertain parameters and sensitive parameters, one sample simulation does not
give a reasonable estimate of the evacuation time. But
how can one reasonably quantify the impact of uncertain parameters such as the number of injured
pedestrians? We tackle this challenge using uncertainty quantification.

\subsubsection{Simulation setup for uncertainty quantification}
\label{sec:uq_setup}

\begin{figure}
    \centering
        \includegraphics[width=0.6\textwidth]{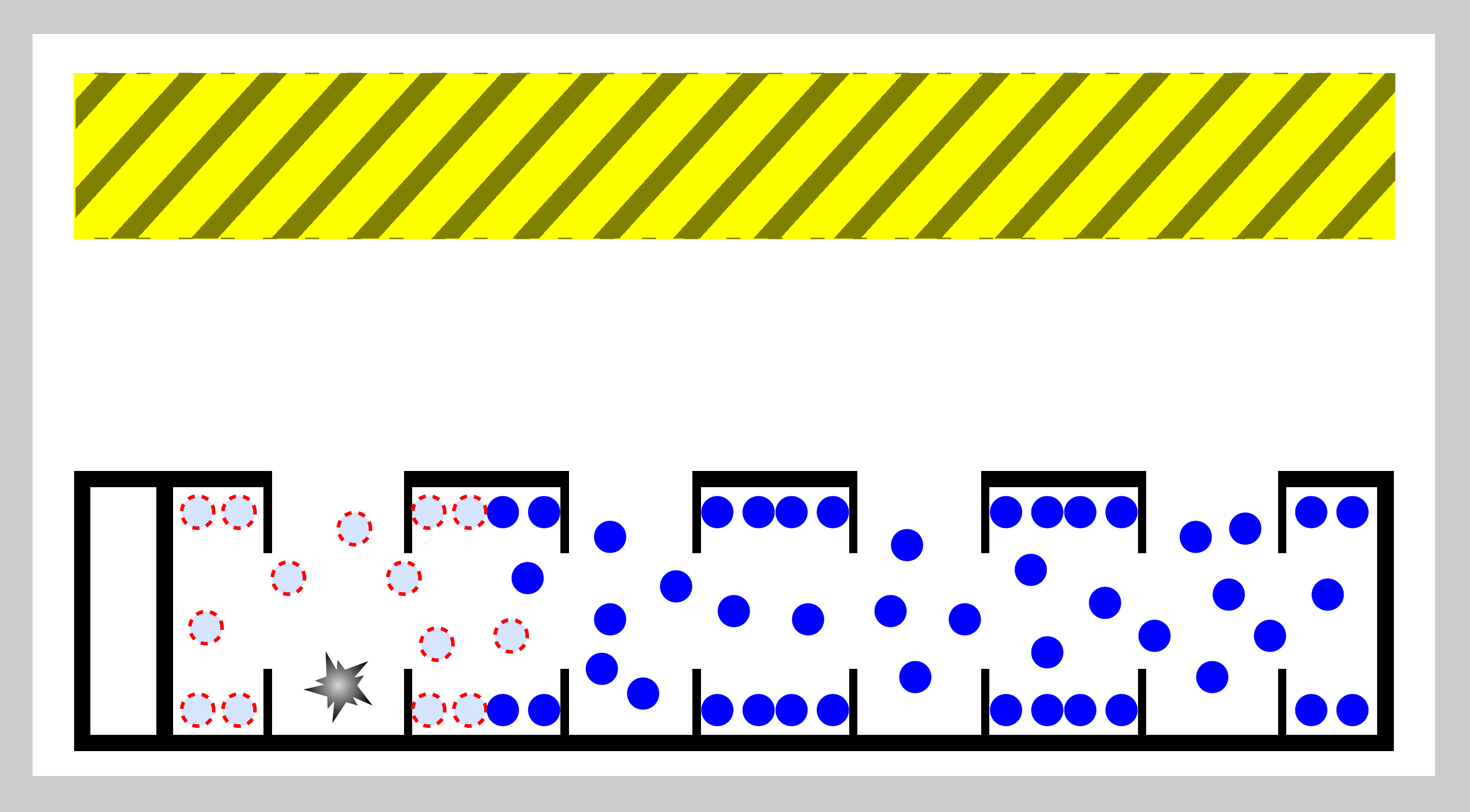}
        \caption{Scenario for the analysis with uncertainty quantification: evacuation of one car with 60 passengers (blue and light blue circles) to a platform (yellow striped rectangle in the upper part of the picture) after a bomb explosion. The grey star shows the place of the bomb. Thus, the red-rimmed light blue circles on the left side of the train are the passengers with a high risk of injury.}
    \label{fig:scenario_startsetting}
\end{figure}

For a proof of concept, we focus on one car and make several assumption with respect
to the initial states of the agents.
Again, every seat is occupied, but this time some people are standing, so that there are 60 persons in the car. They  
evacuate to a `safe' platform next to the car. We assume that the bomb detonated at one end of the
car and that the 14 people near the event are likely to be injured. As long as the
percentage of injured pedestrians is set below 25\%, the attribute of being injured is randomly
assigned to these pedestrians. Above 25\%, all of 14 likely casualties are
marked as injured. For the remaining number of casualties other passengers are randomly chosen.
The percentage of passengers sharing a social $perc_{injPeds}$ is completely unknown. 
It is the first uncertain parameter we investigate. The attribute is randomly assigned to passengers using a uniform distribution according to $perc_{injPeds}$. Figure~\ref{fig:scenario_startsetting} illustrates the setting.

Further uncertain parameters are the percentage of casualties and the speed at which an aide and charge evacuate.
Since there is no data on parameter distributions from real evacuations or experiments, we need to make plausible assumptions. We assume that the parameters are uniformly distributed with the minimum and maximum values in Table \ref{tbl:parametersSocialUQ}.

\begin{table}
     \centering
        \begin{tabularx}{0.95\textwidth}{|p{0.2\textwidth}|X|c|c|}
            \hline
            Parameter              & Description                                  & Min. & Max.\\
            \hline
            $perc_{sharingSI}$     & people sharing a social identity             & 0.6  & 1.0\\
            $perc_{injPeds}$       & percentage of injured pedestrians            & 0.1  & 0.3\\
            $v_{inj}$              & speed of a helper with an injured pedestrian & 0.4  & 0.8\\
            \hline
        \end{tabularx}
        \caption{Three uncertain parameters in the Social Identity Model Application. They are uniformly distributed between their minimum and maximum values.}
     \label{tbl:parametersSocialUQ}
\end{table}

We choose the maximum evacuation time and the number of people who have not yet reached safety 
as quantities of interest. At
each time step, we analyse the mean and the variance or standard deviation 
of each quantity of interest, until all persons have evacuated.

The next step is to calculate percentiles which measure which percentage of the
determined values are below a specific value. By plotting values of the 10th percentile
and the 90th percentile, the space between this values can be interpreted as the area where 80\% of the
values lie.

\subsubsection{Impact of the uncertain parameters}
\label{sec:uq_results}

In real world scenarios, all parameters are uncertain at the same time. In simulations one must focus on the parameters that one expects to have a decisive impact on
 simulation results.
 Here we select the number of pedestrians with social identification $perc_{sharingSI}$, the number of injured
pedestrians $perc_{injPeds}$, and the evacuation speed $v_{inj}$. In this first
uncertainty investigation, all three parameters are assumed to be uniformly distributed as listed in Table
\ref{tbl:parametersSocialUQ}. 
   
\begin{figure}
    \centering
        \includegraphics[width=\textwidth]{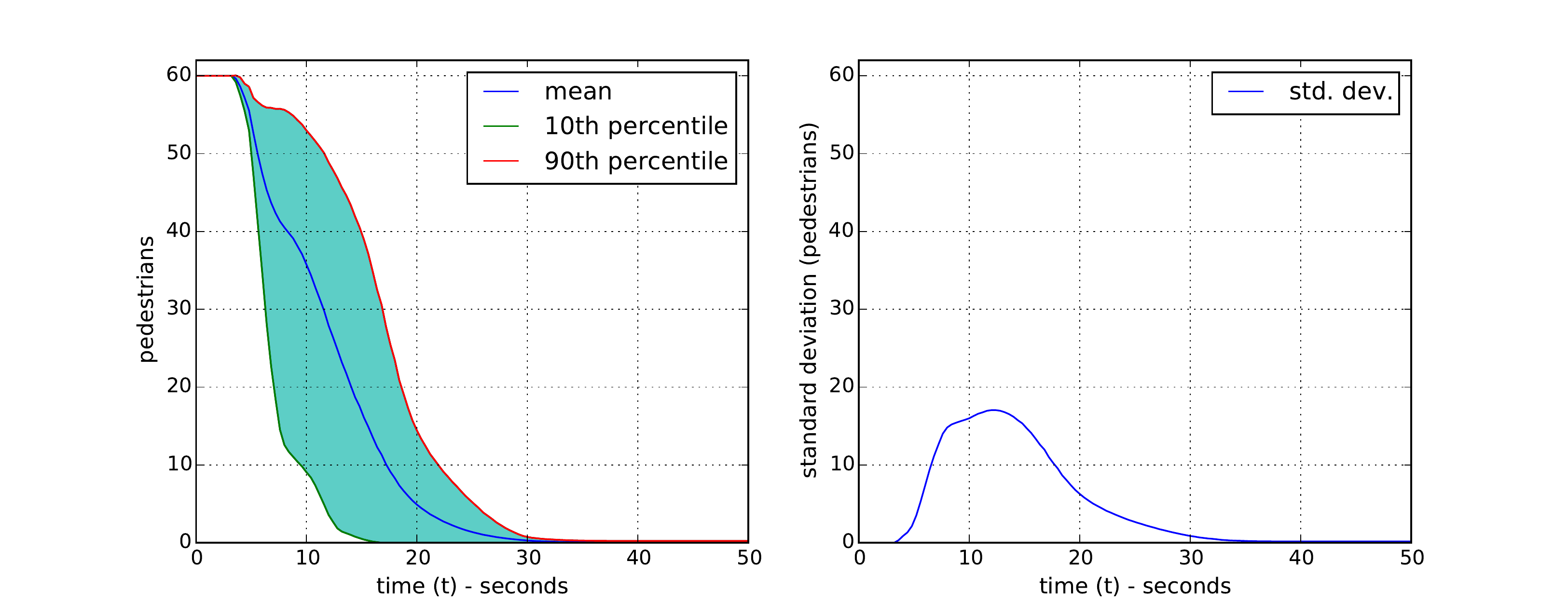} 
        \caption{Uncertainty quantification with  three uniformly
        distributed uncertain parameters. The mean value and the percentiles of the number of pedestrians who remain in the danger zone
				are plotted on the left. The standard deviation is plotted on the right.
				In the simulations, the percentage of pedestrians with social identification $perc_{sharingSI}$ 
				varies between 60\% and 100\%, the percentage
        of injured pedestrians $perc_{injPeds}$ ranges from 10\% to 30\%, and the speed of a helper
        with charge ($v_{inj}$) ranges from $0.4$ m/s to $0.8$ m/s.
        }
    \label{fig:uq_sc_uniform_all_po_005_co_020}
\end{figure}

The quantification results are shown in Figure \ref{fig:uq_sc_uniform_all_po_005_co_020}. In the left
plot, the blue line in the filled space is the mean value of the number of pedestrians remaining in the car.
Values are plotted from the beginning of the evacuation to the end. Therefore, at the beginning all pedestrians are
still in the danger zone, while at the end, everyone has left. The green line on the left of the filled space is the 10th
percentile. The red line on the right of the filled space is the 90th percentile. The filled space
can be interpreted as follows: With a probability of 80\%, the
corresponding number of pedestrians remain in the danger zone within the
time span. During the very first seconds every agent is evacuating, but nobody has successfully evacuated, because it takes
some time to reach the safe area. In the right plot of Figure
\ref{fig:uq_sc_uniform_all_po_005_co_020}, the standard deviation for the number of pedestrians who are still in danger is
plotted. It illustrates the spread around the mean number of agents at every time step
from the beginning of the simulation to the end.
The mean for the maximum evacuation time is 21.35 seconds. The  standard deviation is 5.68 seconds (32.37 variance).

To analyse the impact of a single parameter, only this parameter is disturbed and the others are
kept fixed. We choose the average of the minimum and the maximum values from Table
\ref{tbl:parametersSocialUQ} as the fixed values.


\begin{figure} 
    \centering
        \includegraphics[width=\textwidth]{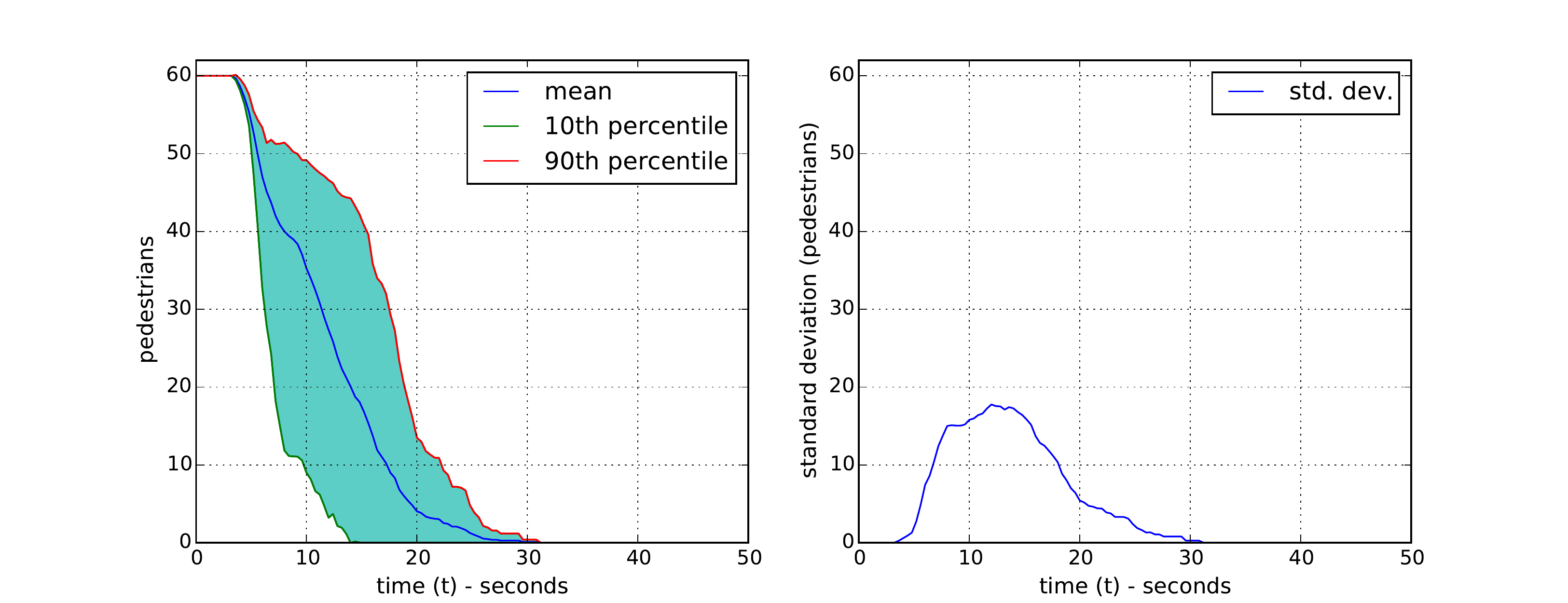}
		\caption{Uncertainty quantification with the percentage of injured pedestrian as uncertain parameter.
		The mean value and the percentiles of the number of pedestrians who remain in danger are plotted on the left. 
        The standard deviation is plotted on the right. Parameter $perc_{injPeds}$ is uniformly distributed between 10\% and 40\%.  
		 }
	\label{fig:uq_sc_injuredPedestrains_min_01_max_03_po_005_co_020}
\end{figure}

Figure \ref{fig:uq_sc_injuredPedestrains_min_01_max_03_po_005_co_020} shows the results where the
percentage of the injured pedestrians $perc_{injPeds}$ is uncertain. Not surprisingly, we observe that the percentage of injured pedestrians has a strong impact on the number of pedestrians who still remain in the danger zone during the whole simulated evacuation time. 
During the first seconds of the evacuation, the spread around the mean is small. This is because
the number of casualties does not have an impact on the behaviour of agents who are not sharing a social identity and
directly evacuate without helping. The mean for the maximum evacuation time is 20.60 seconds and the standard deviation is 5.43 seconds.

\begin{figure}
	\centering
		\includegraphics[width=\textwidth]{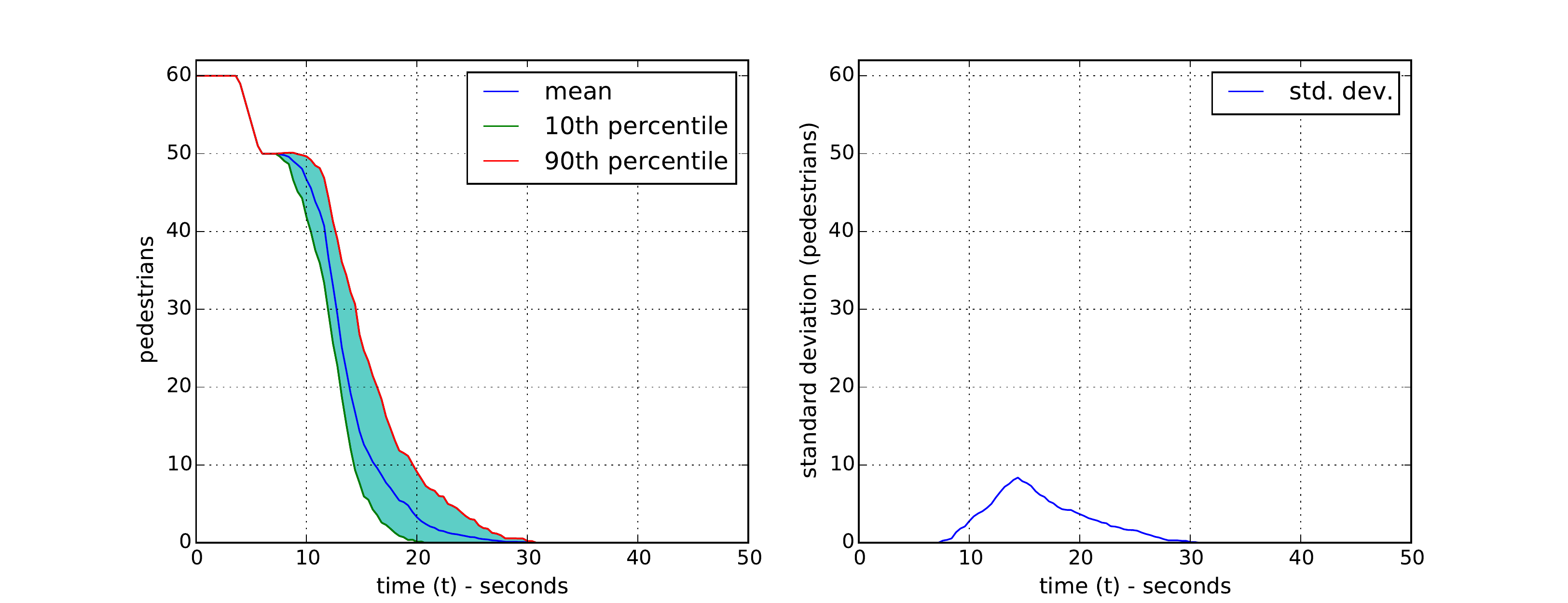}
		\caption{Uncertainty quantification  with the speed of a helper with charge as uncertain parameter.
		The mean value and the percentiles of the number of pedestrians who remain in danger are plotted on the left. 
        The standard deviation is plotted on the right. Parameter
		$v_{inj}$ is uniformly distributed between 0.1 m/s  and 0.4 m/s.
		 }
	\label{fig:uq_sc_meanInjuredSpeed_min_04_max_08_po_005_co_020}
\end{figure}

Next we consider the speed of a helper with an injured pedestrian $v_{inj}$ as uncertain. 
The result is depicted in Figure~\ref{fig:uq_sc_meanInjuredSpeed_min_04_max_08_po_005_co_020}.
During the first seconds, there is no uncertainty in the number of safely evacuated pedestrians. This
can be explained with the immediate evacuation of the unharmed pedestrians who do not share a
social identity and who are near the doors. For these pedestrians the speed of a helper with charge plays no role.
The mean for the maximum evacuation time is 21.81 seconds and the standard deviation is
3.29 seconds.  

\begin{figure}
	\centering
		\includegraphics[width=\textwidth]{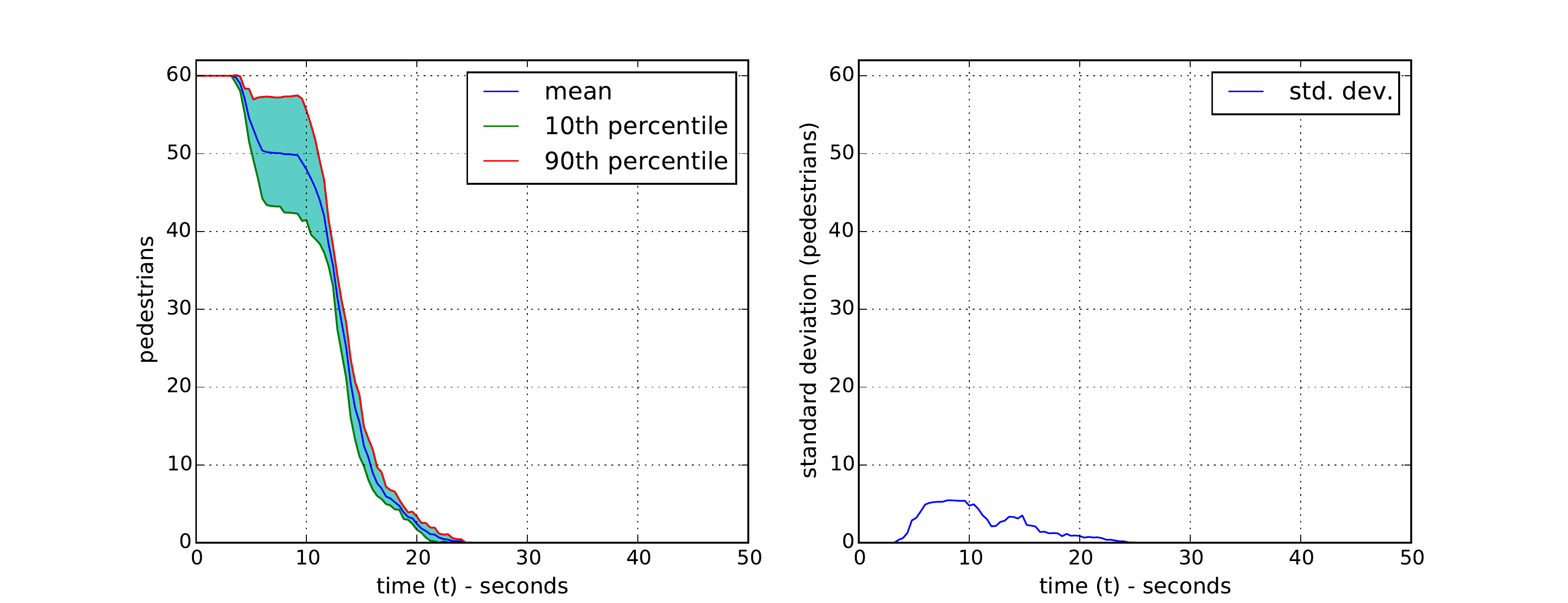}
		\caption{Uncertainty quantification with the percentage of pedestrians sharing a
		social identity as uncertain parameter. 
		The mean value and the percentiles of the number of pedestrians who remain in danger are plotted on the left. 
        The standard deviation is plotted on the right. Parameter
		$perc_{sharingSI}$ is uniformly distributed between 60\% and 100\%.
		}
	\label{fig:uq_sc_sharingIdentity_min_06_max_10_po_005_co_020}
\end{figure}

Finally, Figure \ref{fig:uq_sc_sharingIdentity_min_06_max_10_po_005_co_020} illustrates the results 
when the percentage of pedestrians with a shared  social identity is uncertain.
At the beginning of the evacuation, the number of aides has a strong influence on the number of pedestrians remaining in the danger zone. We find this plausible, because many potential helpers make it likely that injured pedestrians are reached
quickly and that helping behaviour has an impact on how the situation evolves. Later in the simulation, the influence of the parameter
decreases as one might expect: As long as enough unharmed pedestrians share a social identity each
injured person will be helped eventually. The values for the maximum evacuation time spread only a
little bit around the mean of 21.62 seconds with a value of 1.11 seconds for the standard deviation.

\begin{table}
     \centering
        \begin{tabularx}{0.95\textwidth}{|X|c|c|c|}
            \hline
              & \multicolumn{2}{c|}{Max evacuation time}\\
                \cline{2-3}
            Uncertain parameter                          & Mean    & Std. dev.\\
            \hline
            all three parameters                         & 21.35 & 5.68\\
            people sharing a social identity             & 21.62 & 1.11\\
            percentage of injured pedestrians            & 20.60 & 5.43\\
            speed of a helper with an injured pedestrian & 21.81 & 3.29\\
            \hline
        \end{tabularx}
        \caption{Comparison of impact of the uncertain  
				parameters in the Social Identity Model Application on the maximum evacuation times.}
     \label{tbl:maxEvacuationTimeUQResult}
\end{table}


Table \ref{tbl:maxEvacuationTimeUQResult} gives a survey of the results of the four uncertainty quantifications above. From the standard deviation, we see that the percentage of
injured pedestrians has the greatest impact on the simulation results. 
In an earlier sensitivity study \citep{davidich-2013} where the evolution of a passenger stream at a German railway station was simulated and compared to video footage, it was found that changing the targets for the pedestrians had the greatest impact on results compared to other parameters like the passengers' preferred walking speeds. Interestingly, this coincides with our new results: In the Social Identity Model Application each injured agent changes, first and above all, targets: for itself and for potential aides. The percentage of pedestrians who share a social identity, on the other hand, has a comparatively
small impact.
Moreover, if all three parameters are uncertain the impact of the different uncertain parameter
does not appear to be cumulative: The standard deviation of 5.68 seconds in the maximum evacuation time is only marginally higher than
the standard deviation of 5.43 seconds when the percentage of injured pedestrians is the
only uncertain parameter.

Through uncertainty quantification we attain a better and deeper understanding of our new model and the parameters -- an understanding that cannot be derived from the model itself but is important to safety scientists. And we have good news: While the number of pedestrians sharing a social identity, or the degree of this identification, defies measurement, at least at present, the impact of its variation seems small.  Thus, the model retains its predictive power. 

Finally, the results quantitatively substantiate our earlier claim that modelling social identity and the helping behaviour that ensues has a very significant impact on evacuation simulations. Neglecting helping behaviour leads to quantitative results at one extreme end of the analysis. In view of the range of the evacuation times in Figure~\ref{fig:uq_sc_injuredPedestrains_min_01_max_03_po_005_co_020}, for example, neglecting injured pedestrians would lead to a serious underestimation of evacuation times.

\section{Conclusions} 

In this paper, we presented an algorithmic formulation of empirical findings from social psychology
on human behaviour in a situation of great danger and duress. In particular, we looked at the effect of social identification 
within a crowd and ensuing helping behaviour during an evacuation.  
For this, we  embedded our algorithm into a pedestrian evacuation simulation. 
We examined the behaviour in the simulation for a particular scenario that resembled the bomb attack on a metro train in London 
on July 7\textsuperscript{th} 2005. 
The computer simulation reproduced observations from the real evacuation. In particular, the agents 
evacuated in pairs of injured passengers and helpers and the overall behaviour was orderly. 
We argue, that this constitutes a qualitative validation of the computer model.

Crucially, we  went a step beyond qualitative validation:
In most cases, one or more model parameters that influence simulation outcomes are uncertain. Either they are entirely unknown, such as the number of injured pedestrians in our virtual evacuation, or they are measured with limited accuracy.  We identified three parameters that could be decisive in our model: the percentage of injured pedestrians, the speed at which helpers and charges evacuate, and the percentage of people who share a social identity.
We used uncertainty quantification to quantify their influence in an example scenario from which more complex scenarios can be derived. Variations in the percentage of injured pedestrians turned out to have a great influence, whereas variations in the speed had a medium impact, and variations in the percentage of pedestrians sharing a social identity had a relatively small impact. Since the latter parameter is very hard to measure, this is encouraging news for the safety scientist who needs predictive power of the model to give safety advise on the basis of simulations. 

Quantitative validation of our model against measurements, such as trajectories of pedestrians or evacuation times, is still open and must remain open until suitable data is available. Knowing this, we reported all model parameters so that independent researchers can replicate and thus validate -- or falsify -- our findings. 

We consider the instantiation of social identity and helping among strangers in our computer model as a proof of concept that it is indeed possible to carry over findings from social psychology into computer models that possess predictive power.

Yet, helping among strangers in emergencies is only one behaviour among many that stem from social
identification and that are relevant to safety science. Another important example is the
identification with ones own family. Also, with competing social identities the question arises
which of the social identities is salient in which situation. Strategies to handle this must be
found. Methods from uncertainty quantification which we have introduced to the field of safety
science, promise to allow efficient characterisation and quantification of the influence of
competing  and interworking identities and of further social phenomena.

\section*{Acknowledgements}
This work was funded by the German Federal Ministry of Education and Research through the projects MEPKA on mathematical characteristics of pedestrian stream models (grant number 17PNT028) and MultikOSi on assistance systems for urban events -- multi criteria integration for openness and safety (grant number 13N12824), and the Engineering and Physical Sciences Research Council (grant number EP/L505109/1). 
The authors also acknowledge the support by the Faculty Graduate Center CeDoSIA of TUM Graduate School at Technische Universit\"{a}t M\"{u}nchen, Germany.

\bibliographystyle{elsarticle/model2-names}
\bibliography{Literature}







\end{document}